\documentclass[showpacs,twocolumn,aps]{revtex4}
\usepackage{epsf}
\begin{document}
\title{Self-Similarity in Random Collision Processes} 
\author{Daniel ben-Avraham$^1$}
\author{Eli Ben-Naim$^2$}
\author{Katja Lindenberg$^3$}
\author{Alexandre Rosas$^3$}
\affiliation{${}^1$Physics Department, Clarkson University, Potsdam NY
13699-5820\\
${}^2$Theoretical Division and
Center for Nonlinear Studies, Los Alamos National Laboratory, Los
Alamos, New Mexico, 87545\\ 
${}^3$Department of Chemistry and Biochemistry, University of
California San Diego, La Jolla, CA 92093}
\begin{abstract}

Kinetics of collision processes with linear mixing rules are
investigated analytically. The velocity distribution becomes
self-similar in the long time limit and the similarity functions have
algebraic or stretched exponential tails. The characteristic exponents
are roots of transcendental equations and vary continuously with the
mixing parameters. In the presence of conservation laws, the velocity
distributions become universal.

\end{abstract}
\pacs{05.40.-a, 05.20.dd, 02.50.Ey}

\maketitle 

Collision processes underlie fundamental phenomena such as heat
transport in gases \cite{rd} and mixing in fluid flows \cite{omtfjk}.
In closed systems, conservation of mass and momentum implies
Maxwellian velocity statistics \cite{jcm}. However, there are physical
systems such as granular media \cite{pl} and atomic collisions
\cite{prb} where energy or even momentum \cite{rbl} are not
conserved. The reason may be that only a subset of the system is
considered, that not all degrees of freedom are measured (2D imaging
of 3D systems), or that acoustic or other excitations are
ignored. Non-Maxwellian velocity statistics are found in granular
gases \cite{rm}, colloids \cite{wclsa}, turbulent flows \cite{lvcab},
and laser cooling \cite{bbac}.

Random collision processes are a framework for studying the role of
conservation laws, and demonstrate how anomalous velocity statistics
emerge when conservation laws are relaxed
\cite{bmp,kb,eb,bc}. Motivated by this, we consider binary collision
processes with {\it arbitrary} linear collision rules. While in the
long-time limit velocity distributions are generically self-similar,
there is a wide spectrum of possible behaviors. The velocity
distributions are characterized by algebraic or stretched exponential
tails and the corresponding exponents depend sensitively on the
collision parameters. Interestingly, when there is energy or momentum
conservation, the behavior is universal.

Let us consider the most general linear collision law: when a particle
of velocity $u_1$ collides with a particle of velocity $u_2$, its
post-collision velocity $v_1$ is given by
\begin{equation}
\label{rule}
v_1=pu_1+qu_2,
\end{equation}
with $p$ and $q$ the mixing parameters. Special cases include elastic
collisions ($p=0$, $q=1$), inelastic collisions ($p+q=1$), the
granules model ($p+q<1$) \cite{rbl}, the Kac Model ($p^2+q^2=1$)
\cite{mk}, inelastic Lorenz gas ($p=0$, $q<1$) \cite{mp}, and addition
($p=q=1$) \cite{tk}.


Further, we consider perfectly random dynamics: two randomly chosen
particles collide according to (\ref{rule}). The normalized velocity
distribution $P(v,t)$ obeys
\begin{eqnarray}
\label{be} 
{\partial\over\partial t}P(v,t)&=& \int\int du_1 du_2
P(u_1,t)\,P(u_2,t)\\
&\times&\left[\delta(v-pu_1-qu_2)-\delta(v-u_1)\right].
\nonumber
\end{eqnarray}
This Boltzmann equation is termed the Maxwell model in kinetic theory
(the constant collision rate is set to unity) \cite{mhe}.  The
quadratic integrand in Eq.~(\ref{be}) reflects the binary nature of
the collision process and the gain term reflects the collision rule
(\ref{rule}). The number density is conserved by Eq.~(\ref{be}), $\int
dv P(v,t)=1$.

The convolution structure of the evolution equation suggests the
Fourier transform \hbox{$F(k,t)=\int dv e^{ikv} P(v,t)$}. This
quantity obeys the nonlinear and nonlocal equation
\begin{equation}
\label{four-eq} 
\frac{\partial}{\partial t}F(k,t)+F(k,t)=F(pk,t)F(qk,t).
\end{equation}
This closed equation is amenable to analytical treatment.  Moments of
the velocity distribution, $M_n(t)=\int dv v^n P(v,t)$, obey a closed
hierarchy of equations \cite{bk}
\begin{equation}
\label{mom-eq} 
\frac{d}{dt}M_n+\lambda_nM_n=\sum_{m=1}^{n-1}{n\choose m}
p^mq^{n-m}M_mM_{n-m}
\end{equation}
with the shorthand notation $\lambda_n=1-p^n-q^n$. These equations are
solved recursively with $M_0(t)=1$.

We are interested in the long time limit and we seek similarity
solutions of the form 
\begin{equation}
P(v,t)\to e^{\alpha t}\Phi(ve^{\alpha t})\;,\quad \text{as\ }t\to\infty\;.
\label{Pscaling}
\end{equation}
Equivalently, the Fourier transform has the similarity form
\hbox{$F(k,t)\to f(ke^{-\alpha t})$}. This function satisfies
\begin{equation}
\label{four-scl-eq} -\alpha zf'(z)+f(z)=f(pz)f(qz).
\end{equation}
The similarity function may include both a regular and a singular
component $f(z)=f_{\rm reg}(z)+f_{\rm sing}(z)$ with $f_{\rm
reg}(z)=\sum_n \frac{(iz)^n}{n!}f_n$. Normalization sets $f_0=1$. The
leading small-$z$ behavior of the singular component, $f_{\rm
sing}(z)\sim z^\nu$, reflects an algebraic tail of the velocity
distribution
\begin{equation}
\label{tail}
\Phi(w)\sim w^{-\nu-1},
\end{equation}
as $w\to \infty$.  Substituting the leading singular behavior
$f(z)-1\sim z^{\nu}$ into the governing equation (\ref{four-scl-eq})
yields a relation between the scaling parameter $\alpha$ and the
mixing parameters $p$ and $q$
\begin{equation}
\label{root}
\alpha=\nu^{-1}\lambda_\nu.
\end{equation}
There are two types of scaling solutions, depending on the average
initial velocity $M_1(0)$: type-I scaling ($M_1(0)\neq 0$) and type-II
scaling ($M_1(0)=0$).

\begin{figure}
\centerline{\epsfxsize=7cm\epsfbox{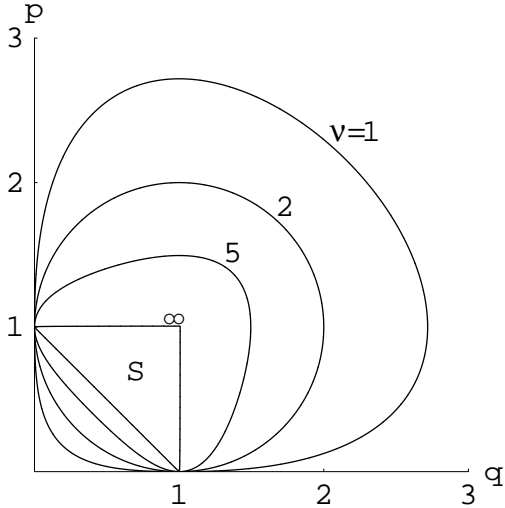}}
\caption{The phase diagram for type-I Scaling. Shown are equi-$\nu$
contours as a function of the mixing parameters.}
\end{figure}

\noindent{\it Type-I Scaling.} The first moment varies exponentially
with time, $M_1=e^{-\lambda_1 t}$, according to the rate equation
$\frac{d}{dt}M_1+\lambda_1M_1=0$ (its initial value can be set to
unity). The small wave-number behavior of the regular component of the
Fourier transform is therefore $F(k,t)\cong 1+ike^{-\lambda_1
t}$. When $\nu>1$, this component dominates over the singular
component. Therefore, $f(z)\cong 1+iz$ and the scaling parameter is
$\alpha=\lambda_1$. When $\nu<1$, the singular component dominates
over the regular one, so the parameter $\alpha$ is not obvious. There
is a spectrum of possible $\alpha$'s depending on $\nu$ according to
the ``dispersion'' curve (\ref{root}). This curve has a maximum
\hbox{$\alpha=\frac{d}{d\nu}\lambda_\nu$} at $\nu$ given by
\hbox{$\nu\frac{d}{d\nu}\lambda_\nu=\lambda_\nu$}. We argue that this
maximum is actually realized by the dynamics. Intuitively, the
selection of the extremum point maximizes the typical wave number
$\propto e^{\alpha t}$. The scaling parameter is therefore
\begin{equation}
\label{alpha-I}
\alpha=\cases{
p^\nu\ln \frac{1}{p}+q^\nu\ln \frac{1}{q}&$\nu\leq1$;\cr 
1-p-q&$\nu\geq 1$.}
\end{equation}
The exponent $\nu$ characterizing the algebraic tails is the root of
the transcendental equation (\ref{root}) with $\alpha$ given by
(\ref{alpha-I}). Explicitly, the equations are
\hbox{$p^\nu\ln\frac{e}{p^\nu}+q^{\nu}\ln\frac{e}{q^\nu}=1$} and
\hbox{$1-p^\nu-q^\nu=\nu(1-p-q)$} for $\nu\leq 1$ and $\nu\geq 1$, respectively
(Fig.~1).

Algebraic tails exist as long as the exponent $\nu$ is finite. The
exponent diverges, $\nu\to\infty$, in the limiting cases $p+q=1$,
$p=1$, and $q=1$, defining the triangular region $S$ (Fig.~1).  In
this domain, the singular component disappears and the scaling
function $f(z)$ is analytic. Moreover, the velocity distribution
$\Phi(w)$ has sharp tails and all of its moments are finite. Outside
the region $S$, the exponent $\nu$ is always finite. It varies
continuously as a function of the mixing parameters and it vanishes,
$\nu\to 0$, when $p\to 0$ or $q\to 0$.

\begin{figure}
\centerline{\epsfxsize=7cm\epsfbox{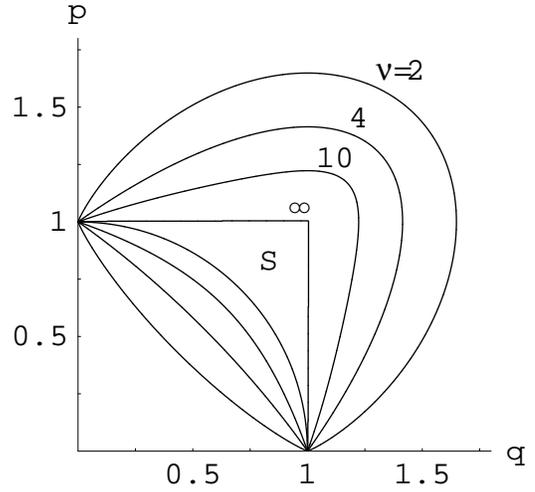}}
\caption{The phase diagram in case II.}
\end{figure}

The curve \hbox{$p\ln\frac{e}{p}+q\ln\frac{e}{q}=1$}, marking the case
$\nu=1$, separates two kinds of behavior. When $\nu>1$ the first
moment characterizes the velocity distribution. In the complementary
case, the typical velocity does not follow from the (integer) moment
behavior. This dichotomy is reminiscent of similarity solutions of the
first and second kind \cite{gib}. Interestingly, extremum selection
determines the typical velocity and thereby the velocity distribution
when $\nu<1$.  Extremum selection similarly governs the speed and the
shape of nonlinear waves in reaction-diffusion problems
\cite{jdm}. Indeed, in terms of the variable $\ln v$, the
similarity solution (\ref{Pscaling}) is nothing but a travelling wave
$\tilde P(\ln v,t)\to \tilde\Phi(\ln v+\alpha t)$.

\noindent{\it Type-II scaling.} When the average initial velocity
vanishes, the initial variance can be set to unity. From the moment
equations (\ref{mom-eq}), the variance varies exponentially with time,
$M_2(t)=e^{-\lambda_2 t}$. The small-wave number behavior $F(k,t)\cong
1-\frac{1}{2}k^2e^{-\lambda_2 t}$ dominates when $\nu>2$ and
consequently, the second moment characterizes the velocity
distribution, $\alpha=\frac{1}{2}\lambda_2$. Otherwise, the singular
component governs the behavior as above and the scaling parameter is
\begin{equation}
\label{alpha-II}
\alpha=\cases{
p^\nu\ln \frac{1}{p}+q^\nu\ln \frac{1}{q}&$\nu\leq 2$;\cr 
\frac{1}{2}(1-p^2-q^2)&$\nu\geq 2$.}
\end{equation}
The exponent $\nu$ characterizing the algebraic tails is the root of
the transcendental equation
\hbox{$p^\nu\ln\frac{e}{p^\nu}+q^{\nu}\ln\frac{e}{q^\nu}=1$} for
$\nu\leq 2$ and \hbox{$1-p^\nu-q^\nu=\frac{\nu}{2}(1-p^2-q^2)$} for
$\nu\geq 2$ (Fig.~2). It diverges in the vicinity of the curves $p=1$,
$q=1$, and $p^2+q^2=1$. Inside this region $S$, the similarity
solution $f(z)$ is regular and the velocity distribution $\Phi(w)$ has
sharp tails. Generally, similarity solutions are symmetric, as
$f(z)=f(-z)$.

\begin{figure}
\centerline{\epsfxsize=7cm\epsfbox{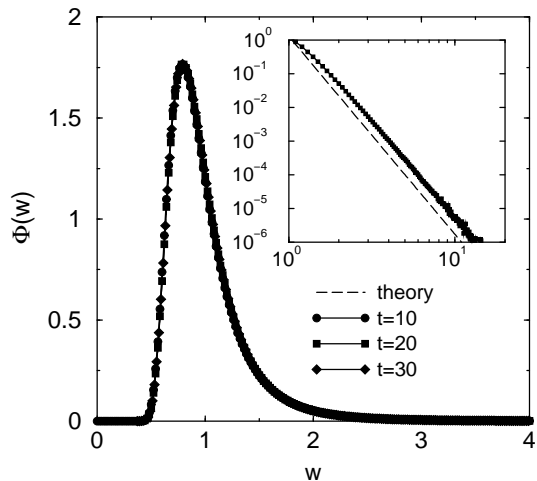}} 
\caption{Self-similarity in type-I scaling. Shown is $\Phi(w)$ versus $w$
for the case $p=q=.4$ ($\alpha=.2$).  In the inset, the tail is
compared with the theoretical prediction $\nu=4.88636$.}
\end{figure}

The diverging exponent $\nu$ indicates that the large velocity tail is
a stretched exponential, rather than algebraic, in the region
$S$. Indeed, the Fourier transform $f(z)\sim \exp(-z^\mu)$ for large
$z$ is compatible with the governing equation (\ref{four-scl-eq}) when
\begin{equation}
\label{root-S}
\lambda_\mu=0. 
\end{equation}
This in turn suggests a stretched exponential behavior 
\begin{equation}
\label{tail-S}
\Phi(w)\sim \exp\left(-w^{\gamma}\right)
\end{equation}
with $\gamma=\frac{\mu}{\mu-1}$ for $w\to\infty$. As was the case for
the exponent $\nu$, $\mu$ is a root of a transcendental equation and
consequently, the exponent $\gamma\geq 1$ varies continuously with the
mixing parameters. Its minimal value $\gamma=1$ is attained along the
region boundaries $p=1$ and $q=1$. The tail is Gaussian, $\gamma=2$,
along the type-II boundary $\lambda_2=0$ and the exponent diverges,
$\gamma\to\infty$, along the type-I boundary $\lambda_1=0$.

Self-similarity holds whether the typical velocity shrinks or grows
with time, i.e., regardless whether $\alpha$ is positive or
negative. For type-I scaling the velocities shrink (grow) with time
when $\lambda_1>0$ ($\lambda_1<0$) and similarly for type-II scaling.

Asymptotically, sufficiently small moments of the velocity
distribution are governed by the typical velocity. Otherwise, the
moment behavior follows from the hierarchy of evolution equations
(\ref{mom-eq})
\begin{equation}
\label{mom}
M_n\sim \cases 
{\exp(-n\alpha t),&$n<\nu$;\cr
\exp(-\lambda_n t),&$n>\nu$.}
\end{equation}
Sufficiently small moments of the velocity distribution exhibit
ordinary scaling behavior while sufficiently large moments exhibit
multiscaling asymptotic behavior \cite{odd}. By multiscaling, we refer
to moment ratios such as $M_n/M_2^{n/2}$ that diverge
asymptotically. 


\begin{figure}
\centerline{\epsfxsize=7cm\epsfbox{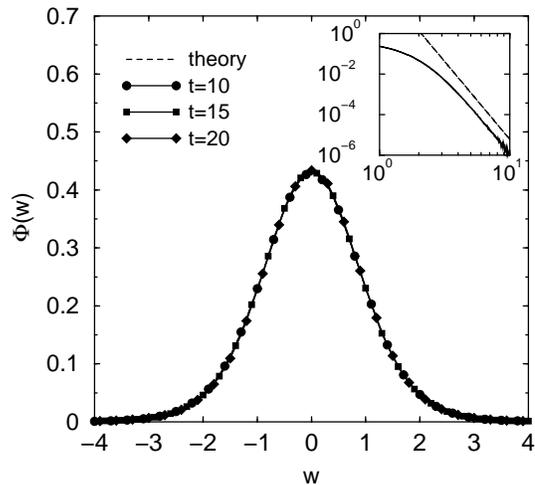}} 
\caption{Self-similarity in type-II scaling. Shown is $\Phi(w)$ versus $w$
for the case $p=q=.6$ ($\alpha=0.14$). The inset compares the tail
with the theoretical prediction $\nu=6.66937$.}
\end{figure}

Numerical simulations confirm the theoretical findings. In the
simulations, randomly chosen pairs of particles undergo the collision
process (\ref{rule}). The number of particles was $10^7$ and the
velocity distributions were obtained from an average over $10$
independent realizations. Simulation results corresponding to flat
initial distributions with support in $[0:1]$ and $[-1:1]$ are shown
in Fig.~3 and Fig.~4, respectively.

There are a number of special cases worth highlighting.

\noindent{\it 1. Addition ($p=q=1$):} This integrable case nicely
demonstrates how similarity solutions emerge.  The transformation
$G(k,t)=1/F(k,t)$ reduces the (local) Ricatti Eq.~(\ref{four-eq}) into
\hbox{$\frac{\partial}{\partial t}G(k,t)+G(k,t)=1$}.  The Fourier
transform reads $F(k,t)=\{1+[F_0^{-1}(k)-1]e^t\}^{-1}$ with
\hbox{$F_0(k)\equiv F(k,t=0)$}. Indeed, the small wave number behavior
of the initial distribution $F_0(k)$ dictates the asymptotic behavior.
When $M_1(0)=1$, type-I scaling occurs, with the similarity solution
$f(z)=[1-iz]^{-1}$ and $\Phi(w)=e^{-w}$ for $w>0$. When $M_1(0)=0$,
type-II scaling occurs, with the similarity solution
$f(z)=[1+\frac{1}{2}z^2]^{-1}$ and
$\Phi(w)=\frac{1}{\sqrt{2}}\exp(-\sqrt{2}|w|)$. One can verify that
these similarity solutions satisfy (\ref{four-scl-eq}) with $\alpha=1$
and $1/2$ for type-I and type-II, respectively.

\noindent{\it 2. Kac Model ($p^2+q^2=1$):} For type-II scaling, energy
is conserved since $\lambda_2=0$. In this case, the velocity
distribution approaches a steady state, $\alpha=0$. The equation (\ref{four-eq})
has the solution $f(z)=\exp(-z^2/2)$ and the velocity distribution is
Maxwellian \hbox{$\Phi(w)=(2\pi)^{-1/2}\exp(-w^2/2)$}. For type-I
scaling, energy is not conserved and the velocity distribution is no
longer universal. However, it still exhibits a Gaussian tail.

\noindent{\it 3. Inelastic Maxwell Model ($p+q=1$):} For inelastic
collisions, the total momentum is conserved, $\lambda_1=0$.  Using the
Galilean transformation $v\to v-M_1(0)$, the initial momentum can be
set to zero and so the behavior is always of type-II. The exponent 
$\nu=3$ is the root of the equation
$\lambda_\nu=\frac{\nu}{2}\lambda_2$.  In this particular case, an 
explicit solution can be found \hbox{$f(z)=(1+|z|)e^{-|z|}$} or
\hbox{$\Phi(w)=\frac{2}{\pi}(1+w^2)^{-2}$} \cite{bmp}.  Interestingly,
when there is a conservation law (either momentum or energy), the
similarity solution is independent of the mixing parameters.

\noindent{\it 4. Lorentz gas ($p=0$):} This case corresponds to
inelastic collisions with massive scatterers. The evolution equation
is linear, \hbox{$\frac{\partial}{\partial
t}P(v,t)+P(v,t)=q^{-1}P(vq^{-1},t)$}.  It is useful to consider the
stochastic process the velocity undergoes $v\to v q\to v
q^2\to\cdots$.  The number of collisions is distributed according to a
Poisson distribution with mean equal to time $t$. Therefore, the
velocity distribution is \cite{bk1}
\begin{equation}
P(v,t)=e^{-t}\sum_{n=0}^\infty {t^n\over n!}\frac{1}{q^n}
P_0\left(\frac{v}{q^n}\right).
\end{equation}
In other words, the variable $\ln v$ is Poisson distributed with mean
equal to $t\ln q$, so a finite number of standard deviations away from
the mean $\ln v$ is Gaussian distributed. Thus, the tail of the
distribution is log-normal, \hbox{$P(v)\sim \exp[-(\ln v)^2/(2t\ln
q)]$}. In the limit $\nu\to 0$, no similarity solutions emerge and all
moments of the velocity distribution exhibit multiscaling,
\hbox{$M_n(t)=M_n(0)\exp(-\lambda_n t)$}. We conclude that the linear
collision process (\ref{rule}) is an effective mixing mechanism. No
matter how small either of the mixing parameter is, eventually, the
binary collision process alters the nature of the velocity
distribution. In other words, nonlinearity provides the mechanism for
the self-similar behavior.

The usual physical particle collision region is associated with $0\leq
p+q\leq 1$ \cite{rbl} and restitution coefficient $0\leq q-p\leq
1$. Other applications may involve values of $p$ and $q$ outside of
this region.  We implicitly assumed that the mixing parameters are
positive ($p,q\geq 0$) but the behavior easily extends to the other
three quadrants in the $p-q$ plane. Consider the first two moments
\hbox{$M_1(t)=M_1(0)\exp(-\lambda_1 t)$} and
\hbox{$M_2(t)=[M_2(0)-cM_1^2(0)]e^{-\lambda_2 t}+c
M_1^2(0)e^{-2\lambda_1 t}$} with
\hbox{$c=(2pq)/(\lambda_2-2\lambda_1)$}. The first moment governs the
second moment ($M_2\sim M_1^2$) when $\lambda_2>2\lambda_1$, i.e., in
the circular domain \hbox{$(p-1)^2+(q-1)^2<1$}, that is entirely
contained in the first quadrant. Thus, type-I scaling occurs only in
the first quadrant. Type-II scaling occurs in the other three
quadrants, regardless of $M_1(0)$.

We tacitly assumed that all moments are finite initially. Consider
initial distributions with a leading small-$k$ behavior of the type
$1-F_0(k)\sim k^{\nu_0}$, competing with $f_{\rm sing}(z)\sim
z^\nu$. The initial conditions govern the asymptotic behavior when
$\nu_0<\nu$ and thus, $\alpha=\nu_0^{-1}\lambda_{\nu_0}$. This
generalization of the previous results applies for both type-I scaling
($\nu_0=1$) and type-II scaling ($\nu_0=2$).

In closing, random and linear mixing results in self-similar velocity
distributions. Nonlinearity is responsible for this scaling and
extremum selection may govern the behavior.  The velocity
distributions have either algebraic or exponential tails, with
nontrivial characteristic exponents.  Every possible algebraic tail
and every faster than exponential decay constitutes the spectrum of
behaviors.  Conservation laws play a crucial role, as the velocity
distribution becomes universal when physical quantities (either energy
or momentum) are conserved.

We thank Paul Krapivsky for useful discussions. This research was
supported by DOE contracts W-7405-ENG-36 (EBN) and DE-FG03-86ER13606 (KL),
and by NSF contract PHY-0140094 (DbA).


\begin{thebibliography}{99}
\bibitem{odd} For type-II scaling, since the limiting velocity distribution
is symmetric, odd moments are sub-dominant and decay slower than
(\ref{mom}).
\end{thebibliography}
\end{document}